# deepFEPS: Deep Learning-Oriented Feature Extraction for Biological Sequences

Hamid Ismail (HI), Ph.D. (Corresponding author)

Department of Computational Data Science and Engineering

North Carolina A&T State University

East Market Street, Greensboro, North Carolina 27411, US.

Email: hismail@ncat.edu

Marwan Bikdash (MB), Ph.D.

Department of Computational Data Science and Engineering

North Carolina A&T State University

East Market Street, Greensboro, North Carolina 27411, US.

## Abstract

**Background:**

Machine- and deep-learning approaches for biological sequences depend critically on transforming raw DNA, RNA, and protein FASTA files into informative numerical representations. However, this process is often fragmented across multiple libraries and preprocessing steps, which creates a barrier for researchers without extensive computational expertise. To address this gap, we developed deepFEPS, an open-source toolkit that unifies state-of-the-art feature extraction methods for sequence data within a single, reproducible workflow.

**Results:**

deepFEPS integrates five families of modern feature extractors —k-mer embeddings (Word2Vec, FastText), document-level embeddings (Doc2Vec), transformer-based encoders (DNABERT, ProtBERT, and ESM2), autoencoder-derived latent features, and graph-based embeddings—into one consistent platform. The system accepts FASTA input via a web interface or command-line tool, exposes key model parameters, and outputs analysis-ready feature matrices (CSV). Each run is accompanied by an automatic quality-control report including sequence counts, dimensionality, sparsity, variance distributions, class balance, and diagnostic visualizations. By consolidating advanced sequence embeddings into one environment, deepFEPS reduces preprocessing overhead, improves reproducibility, and shortens the path from raw sequences to downstream machine- and deep-learning applications.

**Conclusions:**

deepFEPS lowers the practical barrier to modern representation learning for bioinformatics, enabling both novice and expert users to generate advanced embeddings for classification, clustering, and predictive modeling. Its unified framework supports exploratory analyses, high-throughput studies, and integration into institutional workflows, while remaining extensible to emerging models and methods. The webserver is accessible at https://hdismail.com/deepfeps2/.

## Background

Extracting numerical features from biological sequences has long been a challenge, largely because these sequences are inherently textual and must be transformed into meaningful numerical representations before they can be used in machine learning. This process often requires considerable programming and computational expertise. The application of machine learning to biological sequences dates back to 1962, when Margaret Oakley Dayhoff and Richard Eck reconstructed evolutionary relationships using a maximum parsimony method, an approach that can be regarded as an early form of computational inference from molecular data [1]. Neural networks entered the field in 1988, when they were first applied to predict the secondary structure of globular proteins [2]. By the 1990s, hidden Markov models (HMMs) had gained popularity for tasks such as gene prediction, protein family classification, and motif discovery [3].

The scope of machine learning in biological sequence analysis expanded dramatically after the completion of the Human Genome Project in 2003 and the subsequent explosion of sequencing data. Algorithms such as Support Vector Machines (SVMs), Random Forests (RF), and kernel methods were increasingly applied to a wide range of problems, including promoter prediction, splice site recognition, noncoding RNA classification, and protein-protein interaction prediction [4, 5]. These traditional machine learning methods typically relied on features derived from classical sequence descriptors, such as nucleotide or amino acid composition, physicochemical properties, and autocorrelation descriptors [6, 7]. Such handcrafted features provided the foundation for training models that could capture sequence patterns and address a variety of supervised learning problems in bioinformatics.

Building on the foundations of traditional machine learning and feature extraction, the advent of deep learning marked a transformative era in biological sequence analysis beginning around 2015. One of the first breakthroughs was DeepBind, which employed convolutional neural networks (CNNs) to predict protein-DNA and protein-RNA binding, showing that deep architectures could surpass classical motif-based approaches [8]. In the same year, DeepSEA extended CNNs to model the chromatin effects of noncoding variants, laying the groundwork for interpreting genome-wide regulatory activity directly from raw sequence [9]. A few years later, AlphaFold (2018–2021) demonstrated the unprecedented capacity of deep neural networks in structural biology, reaching near-experimental accuracy in protein structure prediction and fundamentally reshaping the field [10]. More recently, advances from natural language processing and representation learning have been extensively adapted to genomics and proteomics. Embedding methods such as Word2Vec and FastText have been applied to biological sequences, treating k-mers as tokens to generate dense distributed representations that capture contextual similarity [11, 12]. Extending beyond word embeddings, Doc2Vec has been utilized to obtain sequence-level representations, allowing entire genes or proteins to be embedded into a continuous space for downstream classification tasks [13]. Unsupervised neural architectures like autoencoders have also been employed to derive latent features from DNA, RNA, and protein sequences, compressing high-dimensional sequence space into lower-dimensional informative representations useful for phenotype prediction and variant prioritization [14]. In parallel, graph-based embeddings such as Graph2Vec and node2vec have been applied to biological networks and sequence-derived graphs, enabling representation of motifs, domains, and higher-order relationships not easily captured by linear models [15, 16]. The most impactful advances, however, have come from transformer-based models, which leverage self-attention to capture long-range dependencies in sequences. DNABERT introduced k-mer–based pretraining to model DNA as a language, demonstrating robust performance across a wide spectrum of genomic tasks [17]. In proteomics, ProtBERT, part of the ProtTrans project, applied the BERT architecture at scale to billions of protein sequences, yielding embeddings that generalize across diverse prediction tasks [18]. Similarly, ESM-2, a large-scale protein language model, has provided state-of-the-art embeddings and atomic-level predictions of protein structures across diverse protein families [19]. Together, these innovations signify the deep learning revolution in bioinformatics, where embedding-based representation learning and large-scale neural architectures now lie at the core of sequence-based discovery.

With these advances, it has become increasingly clear that bioinformatics needs tools capable of bridging the gap between classical feature engineering and modern representation learning. As a successor to the original FEPS toolkit [20, 7], which has supported users since 2016 with traditional feature extraction methods, we now introduce deepFEPS, a comprehensive platform that consolidates five families of state-of-the-art embedding-based feature extractors for DNA, RNA, and protein sequences. Unlike conventional approaches, deepFEPS integrates language-model-inspired techniques to transform raw biological sequences into rich numerical representations suitable not only for deep learning but also for traditional machine learning frameworks. Its modules encompass Word2Vec, FastText, and Doc2Vec embeddings, transformer-based encoders, autoencoder-derived latent features, and graph-based sequence embeddings. By moving beyond simple k-mer counts and handcrafted descriptors, deepFEPS captures higher-order dependencies and complex sequence relationships, enabling more powerful and generalizable analysis across a wide range of sequence-based bioinformatics applications.

## Implementation

deepFEPS is implemented as a modular toolkit that consolidates diverse sequence representation methods within a single, reproducible framework. The system is designed to handle DNA, RNA, and protein FASTA inputs through both a web server and a command-line interface, ensuring accessibility for users with different computational backgrounds. At its core, deepFEPS organizes five families of feature extractors, ranging from classical embedding models to modern transformer architectures, into configurable modules that follow consistent input/output conventions. Each module exposes its key parameters, generates analysis-ready feature matrices in CSV format, and produces automated quality-control reports with visual diagnostics. This unified design allows users to seamlessly compare alternative feature extraction strategies while maintaining reproducibility and interoperability with downstream machine- and deep-learning workflows

### Feature extractors

#### 1. k-mer Word2Vec/FastText (sequence token embeddings)

In this approach, biological sequences are first segmented into overlapping k-mers, which serve as tokens. A word embedding model, such as skip-gram or CBOW in Word2Vec [21] or the sub-word-aware FastText [22], is then either trained from scratch or initialized from pretrained weights. The resulting token embeddings are aggregated using pooling strategies such as mean, mean+max, or CLS embeddings (from transformer-based models), yielding a fixed-length vector representation for each sequence. These embeddings preserve the local compositional context of sequences and provide robust baselines for many downstream tasks. Users can tune key parameters such as k-mer size (k), context window, embedding dimensionality, training epochs, negative sampling strategy, and the choice between pretraining or on-the-fly training to optimize performance.

Word and sub-word embeddings have already demonstrated success in several areas of bioinformatics. For example, k-mer–based Word2Vec models have been used to classify protein families and predict subcellular localization with high accuracy [11]. In genomics, embeddings have enabled promoter and enhancer prediction, as well as the identification of splice sites and other regulatory elements [23]. These successes highlight how embedding-based representations can serve as powerful and versatile building blocks for biological sequence modeling.

## 2. Doc2Vec (document embeddings for sequences)

Doc2Vec [24] extends the idea of word embeddings by treating each biological sequence as a "document" and learning a sequence-level embedding jointly with its constituent k-mer tokens. Unlike simple k-mer averaging, this method captures both global context and local composition, making it particularly effective for identifying long-range motifs or integrating compositional and positional information within a sequence. Performance depends on parameters such as the k-mer size (k), context window, embedding dimensionality, number of training epochs, and the choice between Distributed Memory (DM) or Distributed Bag of Words (DBOW) training modes.

Doc2Vec has shown strong potential in bioinformatics applications where understanding the entire sequence context is crucial. For instance, it has been used to predict protein functions and classify protein families, leveraging global representations to distinguish subtle functional differences [13]. In genomics, Doc2Vec embeddings have been applied to enhancer and promoter prediction and to capture regulatory sequence activity [25]. These applications demonstrate that Doc2Vec provides a powerful complement to word-level embeddings, especially in tasks that benefit from holistic sequence representation.

## 3. Autoencoder-derived features

In this approach, biological sequences are first converted into numerical forms, such as k-mer bag-of-words profiles or one-hot encoded windows and then passed through a feed-forward autoencoder. Instead of keeping all the raw input dimensions, the model learns to compress the information into a low-dimensional bottleneck layer, and these activations become the exported features. This compression not only reduces redundancy but also acts as a natural denoiser, emphasizing the most informative aspects of the data. Users can fine-tune the model by adjusting the latent dimensionality, layer widths, activation functions, and regularization strategies such as dropout or normalization, along with training settings like epochs and batch size. Because the mapping is learned directly from the provided dataset (or from a specific corpus) the resulting features are context-aware and tailored to the biological problem at hand.

Autoencoders have already shown promise across many areas of bioinformatics. For instance, they have been used to denoise single-cell RNA-seq data, improving downstream clustering and differential expression analysis [26]. In proteomics, autoencoder-derived features have helped in protein function prediction and subcellular localization [27]. They have also been applied in variant effect prediction, learning compressed sequence features that generalize across species and contexts [28]. These successes highlight the flexibility of autoencoders in uncovering complex biological patterns that traditional handcrafted features may overlook.

## 4. Graph-based embeddings

Graph-based embeddings approach sequence modeling by first representing biological data as a graph; for example, building edges from k-mer co-occurrence within sequences or from sequence-similarity relationships across datasets. Once the graph is constructed, node or graph embeddings can be learned using algorithms such as node2vec-style random walks, which capture relational structure that complements the linear token order. These node embeddings are then aggregated across a sequence (using operations like mean pooling or attention-based weighting) to produce a fixed-length descriptor suitable for downstream tasks. Key parameters include the definition of edges or windows, walk length, number of random walks, and the embedding dimensionality.

Graph embeddings have found increasing success in bioinformatics, particularly where relational dependencies are as important as compositional ones. For example, k-mer co-occurrence graphs combined with node2vec embeddings have been used for viral genome classification and host-pathogen interaction prediction [29, 30]. In proteomics, graph-based methods have proven powerful for protein-protein interaction (PPI) prediction and functional annotation, since they naturally capture network relationships among biomolecules [31]. These examples illustrate that graph embeddings enrich sequence representation by uncovering structural and contextual information often missed by linear models.

## 5. Transformer-based encoders (pretrained protein and nucleotide models)

Transformer-based encoders represent the newest generation of sequence embedding methods, drawing on architectures originally developed for natural language processing. For proteins, models such as ESM-2 provide powerful pretrained embeddings, while for DNA and RNA, k-merized transformer models like DNABERT and its variants have become widely used. In practice, sequences are tokenized using the model's native scheme, with built-in safeguards for long inputs, such as safe truncation or chunking on the server side, to ensure scalability. The resulting embeddings can be aggregated with configurable pooling strategies (CLS, mean, or mean+max) to produce a per-sequence vector representation. By leveraging transfer learning from massive unsupervised sequence corpora, transformer encoders often deliver the strongest out-of-the-box performance across a wide range of functional prediction tasks.

Transformers have already achieved notable success in bioinformatics. ESM models have been applied to protein structure prediction, mutation effect analysis, and function annotation, often rivaling experimental pipelines in speed and accuracy [19, 32]. In genomics, DNABERT has been used for tasks such as promoter and enhancer prediction, splice site recognition, and chromatin state classification, where its ability to model long-range dependencies in DNA provides a major advantage [17]. Recent studies have further demonstrated the versatility of transformer-based language models in proteomics, particularly for post-translational modification (PTM) prediction. For instance, LMPhosSite applied embeddings from local sequence windows alongside a pretrained protein language model to accurately predict phosphorylation sites [33]. Building on this, researchers have shown that integrating embeddings from multiple protein language models can significantly improve the prediction of O-GlcNAc modification sites, highlighting the benefit of ensemble representation learning [34]. Similarly, LMCrot leveraged interpretable transformer-based embeddings to enhance crotonylation site prediction [35], while CaLMPhosKAN introduced a hybrid approach combining codon-aware and amino-acid-aware embeddings with wavelet-based

Kolmogorov-Arnold networks, achieving state-of-the-art performance for phosphorylation site prediction [36]. Together, these works illustrate how transformer-based protein models are not only powerful for structural biology and general function prediction but are also enabling specialized, high-accuracy predictors for PTMs and regulatory biology.

## Results

### Input and preprocessing

deepFEPS accepts multi-FASTA inputs for DNA, RNA, or protein sequences. To prepare data for supervised modeling, the FASTA definition line can be formatted as:

>ID|label

This ensures that the sequence ID is recorded in the first column of the output CSV file, while the label appears in the last column, making the dataset immediately suitable for supervised modeling. Additional options are available depending on the chosen feature extractor.

The web server is designed for small to moderate datasets, making it ideal for quick tests or exploratory analyses. For larger datasets and more advanced workflows, we recommend downloading deepFEPS and running it via the command-line interface (CLI), which supports batch processing and more flexible configurations. Detailed hardware and software requirements for local installation are provided in the GitHub repository.

### Output and automatic QC

Each run produces an analysis-ready CSV in which rows correspond to sequences and columns to features, with optional identifier and label fields when supplied. To make the output immediately interpretable, the web interface provides a quality-control dashboard, as shown in **Figure 2**, that reports the number of sequences processed, the feature dimensionality, matrix size on disk, missing-value counts (imputed as zeros), and overall sparsity; key descriptors for checking whether the dataset is of the expected scale. A lightweight preview table displays the first few rows and capped columns of the feature matrix, allowing users to quickly confirm that identifiers are in the right place, numeric ranges look reasonable, constant columns are absent, and no label information has leaked into the feature set.

Beyond the preview, diagnostic visualizations provide deeper insights into data quality. The row-norm distribution (**Figure 3**) shows the L2 magnitude of each sequence's feature vector. A tight cluster around a central peak suggests consistent scaling across samples, whereas a long right tail signals that some sequences dominate with unusually large values, and a heavy left tail or isolated low bars can reveal near-empty vectors from filtering or tokenization mismatches. Corrective steps include per-row normalization or clipping. The feature-variance distribution (**Figure 4**) summarizes variability across sequences for each feature. Features with near-zero variance contribute little to discrimination and can be removed, while those with very high variance may highlight biologically informative heterogeneity or, in some cases, artifacts of extractor settings. Users can decide whether to apply variance filtering or dimensionality reduction methods such as PCA based on this view.

When labels are present, the class distribution pie chart (**Figure 5**) and table provide a quick check on balance. Equal slice sizes suggest balanced data, while skewed proportions reveal class imbalance that may bias models toward the majority class. In such cases, strategies like class weighting, resampling, or stratified evaluation are recommended. Finally, the heatmap preview (**Figure 6**) offers a visual snapshot of a small slice of the feature matrix after normalization. Vertical bands of uniform color indicate low-variance or constant features, horizontal bands suggest duplicate or highly similar sequences, and block-like patterns reveal correlated groups of features or motif signals. Users can use these patterns to decide whether additional preprocessing, such as feature pruning, deduplication, or correlation filtering, is needed.

Together, these summaries and visualizations do more than describe the dataset: they actively guide users in diagnosing scaling issues, identifying redundant or noisy features, checking label distributions, and spotting structural patterns. By interpreting these cues, users can make principled choices about normalization, feature selection, class balancing, and dimensionality reduction, ensuring more reliable and stable downstream modeling.

**Web Server and Command-Line availability**

deepFEPS is available as both an interactive web server (https://hdismail.com/deepfeps2/) (see **Figure 1**) and a command-line interface (CLI) for scripted, large-scale, or high-throughput workflows. The web server is designed for exploratory use and parameter tuning on small to medium datasets, whereas the CLI exposes advanced options, supports CPU/GPU execution where available, and integrates smoothly with institutional pipelines and workflow managers. Source code and the CLI are hosted at https://github.com/hamiddi/deepFEPS. The web service accepts compressed or plain FASTA uploads, returns a download link for the resulting CSV feature matrix, and automatically records a parameter manifest to facilitate reproducibility.

**Performance considerations**

Computational requirements scale with the choice of extractor and the size of the corpus. Transformer encoders typically incur the highest runtime and memory demand because of large model weights and tokenization overhead, but they often deliver the strongest transfer-learning baselines. Autoencoders and k-mer token embeddings are substantially lighter and easier to scale horizontally across CPUs or modest GPUs. For graph-based embeddings, cost is dominated by graph density and random-walk settings (e.g., window length and number of walks), which trade runtime against representation fidelity. For very large or repeated submissions (and when fine control over hardware, catching pretrained models, batching, or mixed precision is desired) the command-line interface is recommended.

**Limitations and future directions**

Despite its breadth, deepFEPS has several limitations. Transformer encoders have fixed positional capacities; processing very long sequences often requires chunking or sliding windows, which can attenuate long-range dependencies. We plan to expose explicit windowing, stride, and aggregation policies (e.g., mean, attention-weighted, or learned pooling) and to report effective context coverage. Graph-based embeddings remain sensitive to graph construction heuristics—including k-mer co-occurrence definitions, edge weighting, and thresholding—and their performance can vary with graph density; incorporating domain-specific priors (e.g., known interaction networks or pathway constraints) is a priority. Finally, the current QC dashboard provides lightweight, practical indicators; forthcoming modules will add formal diagnostics such as mutual information with labels, redundancy and collinearity measures, stability across resamples, and drift detection, together with exportable reports. Additional roadmap items include enhanced memory/performance profiling, deterministic execution for strict reproducibility, and broader support for pretrained models and offline caching.

## Discussion

deepFEPS builds on and extends earlier sequence feature extraction frameworks. Classical tools such as iFeature [6] and dna2vec [23] provided access to predefined descriptors or word embedding-based representations, but were limited in scope and flexibility. Our earlier toolkit, FEPS [7, 20], offered a webserver and software package for extracting handcrafted protein sequence descriptors, supporting researchers with traditional machine-learning workflows for nearly a decade. While FEPS lowered barriers to protein sequence analysis, it was designed around classical feature engineering approaches.

In contrast, deepFEPS can be viewed as the next-generation successor to FEPS, consolidating five families of modern embedding methods—including k-mer embeddings, document-level models, autoencoders, graph embeddings, and transformer-based encoders—into a single, reproducible workflow. By bridging traditional descriptors with state-of-the-art representation learning, deepFEPS unifies both paradigms in one environment and removes the need to switch between multiple libraries or fragmented pipelines.

The platform is broadly applicable across sequence analysis tasks. By lowering the barrier to advanced embeddings, deepFEPS enables promoter and enhancer prediction, protein function classification, noncoding RNA characterization, and post-translational modification site prediction. It also supports large-scale applications such as genome annotation and protein family clustering, where reproducible, high-quality embeddings are essential for downstream machine- and deep-learning models.

Although we did not include a full benchmarking study here, preliminary applications on protein family datasets showed that transformer-based embeddings captured richer discriminative features compared to k-mer baselines, consistent with recent results from large protein language models [19]. These findings suggest that deepFEPS not only streamlines access to advanced embeddings but also empowers researchers to evaluate, within a single platform, which feature extraction strategy best aligns with their biological question.

## Conclusions

deepFEPS brings modern sequence representation learning into a single, coherent tool that is equally effective as an interactive web service and as a scriptable command-line system. By unifying diverse extractors under consistent input/output conventions, producing reproducible, timestamped outputs, and providing immediate quality-control diagnostics, it shortens the path from raw DNA, RNA, and protein sequences to analysis-ready feature matrices. As the next-generation successor to the FEPS toolkit, deepFEPS lowers the practical barrier to advanced embeddings and supports routine bioinformatics, high-throughput screening, and exploratory computational biology. Its extensible design ensures continued relevance as new representation learning methods emerge, making it a valuable resource for both experimental and computational researchers. The platform is freely available as a web server at https://hdismail.com/deepfeps2/ and as an open-source command-line package at https://github.com/hamiddi/deepFEPS.

## Availability and requirements

- **Project name:** deepFEPS
- **Project home page:** https://hdismail.com/deepfeps2/ (web server)
- **Source code repository:** https://github.com/hamiddi/deepFEPS
- **Operating system(s):** Platform independent (tested on Linux, macOS, and Windows)
- **Programming language:** Python (≥3.8)
- **Other requirements:**
  - Dependencies managed via requirements.txt or environment.yml (e.g., PyTorch ≥1.11, NumPy, scikit-learn, gensim, transformers, networkx, matplotlib)
  - For transformer-based models, GPU acceleration (CUDA-enabled device) is recommended but not required
- **License:** GNU General Public License v3.0 (GPL-3.0)
- **Any restrictions to use by non-academics:** None (open-source and freely available to all users)

## List of abbreviations

- BERT: Bidirectional Encoder Representations from Transformers
- CLI: Command-Line Interface
- CNN: Convolutional Neural Network
- CSV: Comma-Separated Values
- DBOW: Distributed Bag of Words (Doc2Vec training mode)
- DM: Distributed Memory (Doc2Vec training mode)
- DNABERT: DNA Bidirectional Encoder Representations from Transformers
- ESM: Evolutionary Scale Modeling (protein transformer family)
- FEPS: Feature Extraction from Protein Sequences
- GPU: Graphics Processing Unit
- HMM: Hidden Markov Model
- ML: Machine Learning
- NLP: Natural Language Processing
- PCA: Principal Component Analysis
- PPI: Protein–Protein Interaction
- PTM: Post-Translational Modification
- QC: Quality Control
- RF: Random Forest
- ROC: Receiver Operating Characteristic
- SVM: Support Vector Machine
- Word2Vec: Word-to-Vector model (distributed embeddings)

## Declarations

**Ethics approval and consent to participate**
Not applicable.

**Consent for publication**
Not applicable.

**Availability of data and materials**
The web server deepFEPS2 is publicly available at https://hdismail.com/deepfeps2/.

The source code and additional resources are available at https://github.com/hamiddi/deepFEPS.

**Competing interests**
The authors declare that they have no competing interests.

**Funding**
No funding was received for this work.

**Authors' contributions**
HI conceived and designed the study, implemented the computational pipeline, and drafted the manuscript. MB contributed to the conceptual framework, provided critical revisions, and supervised the research process. Both authors read and approved the final version of the manuscript.

**Acknowledgements**
The authors thank colleagues and collaborators at North Carolina A&T State University and BioAGTC for their support and valuable feedback during the preparation of this work.

**Authors' information**

Hamid Ismail (HI), Ph.D., is the Administrator of the Research High-Performance Computing Center and a Lecturer at North Carolina A&T State University (ORCID: 0000-0002-2690-5655)

Marwan Bikdash (MB), Ph.D., is a Professor and Director of the Department of Computational Data Science and Engineering at North Carolina A&T State University.

**Figure 1.** Screenshot of the deepFEPS web server interface displaying the five groups of feature extractors. Each group corresponds to a family of methods (k-mer embeddings, document-level embeddings, autoencoder-derived features, graph-based embeddings, and transformer-based models) allowing users to select and configure pipelines tailored to DNA, RNA, or protein sequences.

**Figure 2.** A summary panel reports the number of sequences, features per sequence, file size, missing-value counts, and sparsity, while a lightweight table displays the first rows and columns

of the feature matrix. Together, these provide a quick check that identifiers, scaling, and basic structure match expectations before downstream modeling.

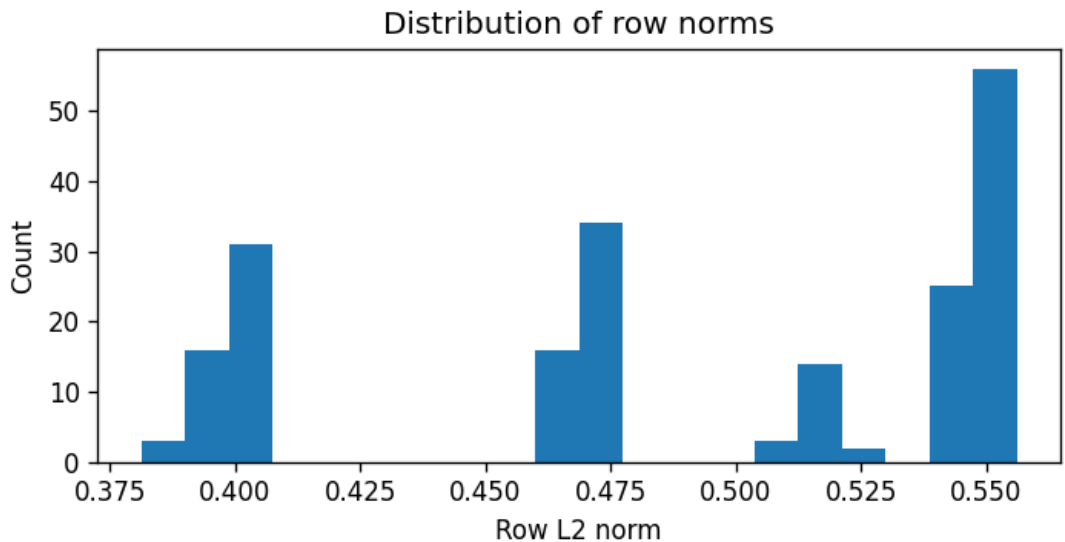

**Figure 3.** Histogram of per-sequence L2 norms showing the overall magnitude of feature vectors. A sharp central peak indicates consistent scaling across samples, while heavy tails or multimodality suggest outliers, heterogeneous inputs, or extractor issues. Small norms may flag near-empty representations; large norms may indicate scaling drift.

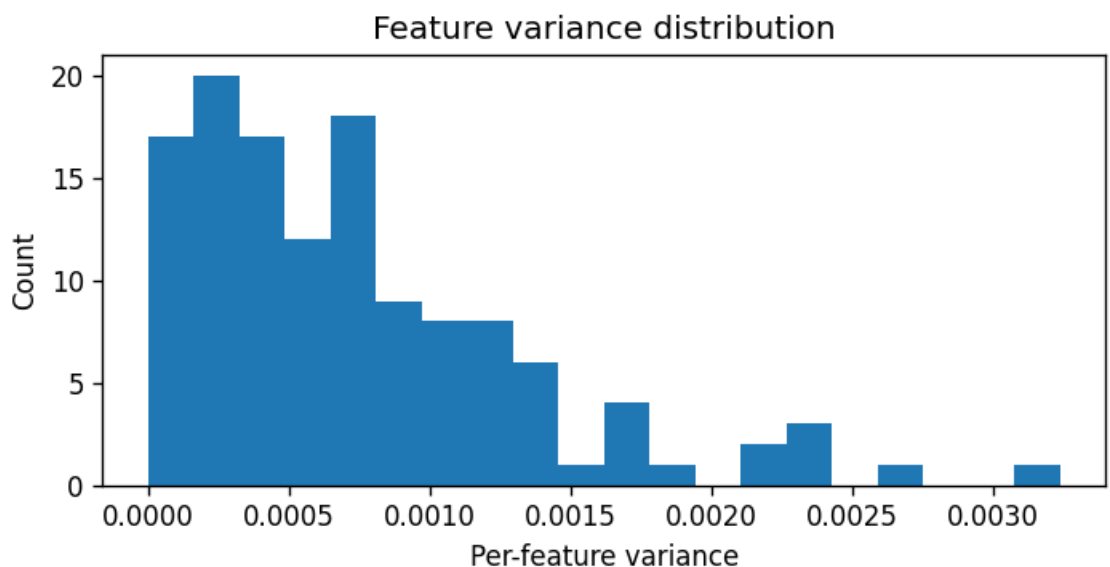

**Figure 4.** Histogram of feature variances across sequences, used to identify low-information or noisy features. Near-zero variance features contribute little and can be removed, while high-variance features may capture biologically relevant heterogeneity or artifacts. The distribution helps guide dimensionality reduction or variance filtering.

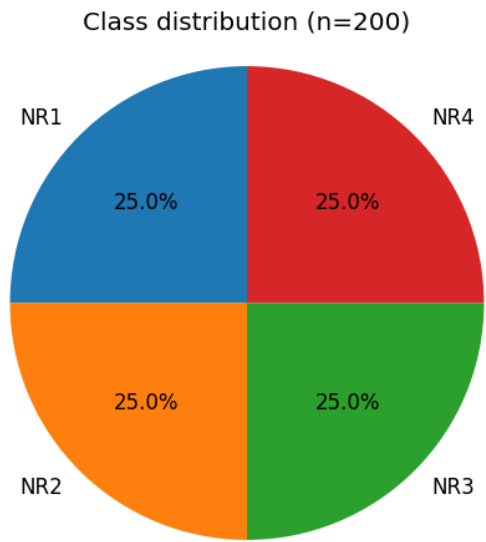

**Figure 5.** Pie chart of class counts when labels are provided. Balanced slices indicate even representation, while skewed proportions highlight imbalance that can bias classifiers and require corrective strategies such as weighting, resampling, or stratified evaluation.

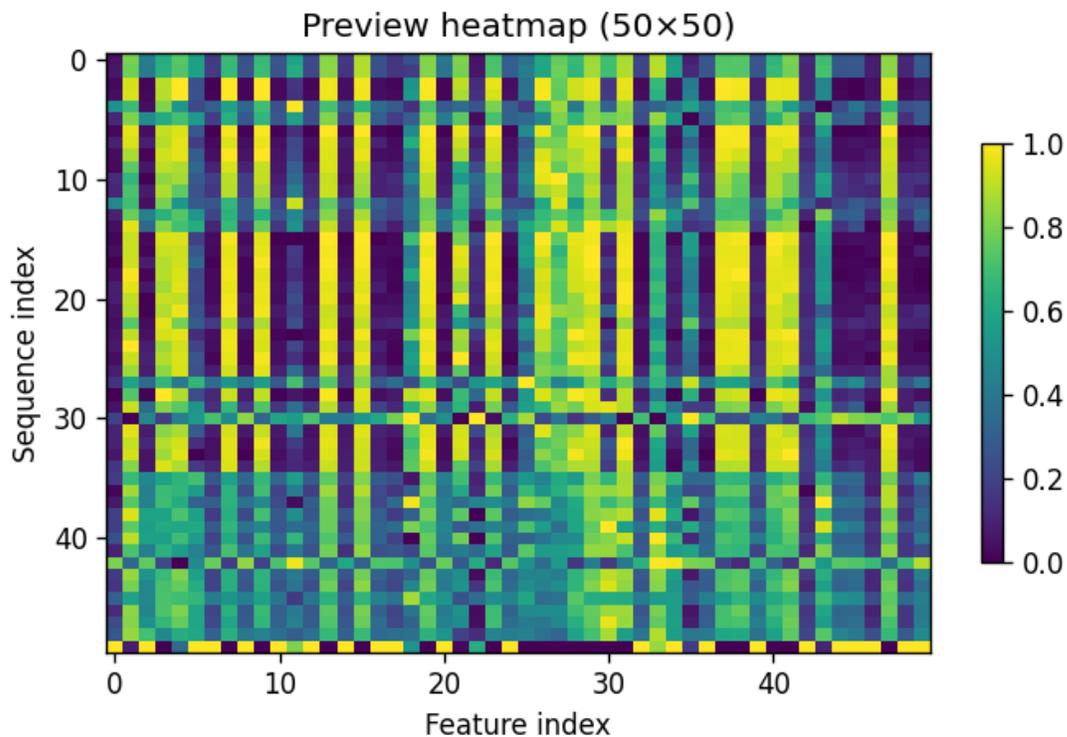

**Figure 6.** Heatmap preview shows normalized visualization of a small block of the feature matrix. Vertical bands indicate constant or low-variance features, horizontal bands suggest duplicated or highly similar sequences, and block-like structures reveal correlated features or motif signals. This preview provides an at-a-glance diagnostic of structural patterns in the data.